# THE ADAPTIVE MEAN-LINKAGE ALGORITHM:
# A BOTTOM-UP HIERARCHICAL CLUSTER TECHNIQUE

HÉLIO M. DE OLIVEIRA

*Departamento de Estatística - Universidade Federal de Pernambuco*
*Cidade Universitária, Recife, PE, Brazil*

E-mail: `hmo@de.ufpe.br`

**Abstract** — In this paper a variant of the classical hierarchical cluster analysis is reported. This agglomerative (bottom-up) cluster technique is referred to as the Adaptive Mean-Linkage Algorithm. It can be interpreted as a linkage algorithm where the value of the threshold is conveniently up-dated at each interaction. The superiority of the adaptive clustering with respect to the average-linkage algorithm follows because it achieves a good compromise on threshold values: Thresholds based on the cut-off distance are sufficiently small to assure the homogeneity and also large enough to guarantee at least a pair of merging sets. This approach is applied to a set of possible substituents in a chemical series.

## 1  Introduction: Fundamentals on Cluster Trees

One approach largely adopted in data analysis and pattern recognition is the cluster analysis [Diday, 1988; Krihnaiah and Kanal, 1982; Arbie et al. 1996; Backer, 1995; Kaufman and Rousseeuw, 1990]. The idea of finding groups in data is normally explored in cluster techniques considering normalised parameters and a metric in order to merge points. For each point $S_i$, a number $p$ of parameters is considered (defining the descriptor space). If the set of descriptor parameters is $\{p_1, p_2, ..., p_p\}$, the corresponding normalised parameters would be

$$X_{ik} = \frac{\pi_k(i) - E(\pi_k(i))}{s.d.(\pi_k(i))} \qquad k=1,2...p, \quad (1)$$

where $X_{ik}$ denotes the value of the $k^{th}$-normalised parameter for the point $S_i$. The expected value $E(.)$ and the standard deviation (s.d.) are calculated over all $i$. The distance between two points $S_i$ and $S_j$ can be computed by means of the Euclidean distance between their respective normalised parameter vectors, that is:

$$d_{ij} = \| X_i - X_j \|. \qquad (2)$$

A matrix of distances between points is then constructed. It is a symmetric matrix with a null main diagonal, henceforth denoted by $(d_{ij})$. The points can be clustered based on such a matrix resulting on a cluster tree or dendrogram. The classical examples of hierarchical clustering appear in biological taxonomy and but several clustering techniques have been adopted in many other areas. There exists divisive (top-down) and agglomerative (bottom-up) hierarchical cluster procedures but this work deals with the second kind of algorithms. Different agglometative hierarchical clustering data description schemes are used depending on the metrics adopted to merge the "nearest" pair of clusters. Some common distance measures lead to the nearest-neighbour algorithm, the furthest-neighbour algorithm, the average-neighbour algorithm and the mean-neighbour algorithm. These algorithms are often associated with an arbitrary "distance threshold" so that clustering terminates when the distance between neighbour exceeds such a threshold [Jain and Dubes, 1988, Diday and Simon 1976]. Such a threshold association yields, respectively, the single-linkage algorithm, the complete-linkage algorithm, the average-linkage algorithm and the mean-linkage algorithm. A new cluster method termed "adaptive mean-linkage algorithm" is introduced in the next section.

## 2  An Adaptive Mean-linkage Algorithm

Linkage algorithms require the choice of a <u>fixed</u> (arbitrary) threshold. The value of such a threshold is rather empirical: it should not be too large or too small. In contrast, the new approach introduces a rational and objective rule to obtain <u>adaptive</u> thresholds based on a minimax criterion. A few definitions and simple results are necessary so as to understand the proposed method. The cut-off distance is defined as the limit distance between points that will be clustered (a threshold). In order to determine the cut-off distance it is required to pick up the lowest value of the Euclidean distance between $S_i$ and any other points $S_j$. This is made for each point $S_i$ and then the greatest obtained value is defined as the cut-off distance, i.e.,

<u>Definition 1</u>. (*Cut-off distance*).
$$d_u = \text{Max}_i \, \text{Min}_j \, d_{ij}. \qquad \blacksquare$$

One point $S_i$ is said to be within the cut-off distance related to the point $S_j$ if and only if the distance $d_{ij}$

between them is less than or equal to the cut-off distance. Denoting by $n_i$ the number of points within the cut-off distance from $S_i$, it is possible to define the following set: A set $\Gamma_i$ is called a $S_i$-neighbourhood if it contains all the $j$-indexes of the points $S_j$ within the cut-off distance to $S_i$, arranged in a non-decreasing order, i.e. $\Gamma_i = \{j_1, j_2...j_{n_i}\}$, in such a way that $0 \leq d_{ij1} \leq d_{ij2} \leq ... \leq d_{ijn_i} \leq d_u$.

Lemma 1 - Each neighbourhood contains at least two distinct elements.
Proof. It is obvious that the point $S_i$ belongs to $\Gamma_i$ so $\Gamma_i$ is not empty. The cut-off distance is also formulated to include at least one closest point. If the point $S_{j*}$ ($j* \neq i$) is the one closest to $S_i$, then $d_{ij*} = \text{Min}_j d_{ij} \leq d_u$, hence $S_{j*}$ also belongs to $\Gamma_i$. ∎

For each point a neighbourhood set is created. From these sets it is possible to create subsets, called sub-neighbourhoods. Let $\Gamma_i^v$ be a sub-neighbourhood from $S_i$, defined as the subset of $\Gamma_i$ that contains just its first $v$ elements. The set $\Gamma_j^v$ denotes a sub-neighbourhood from $S_j$.

Definition 2. (Extremely close points). The points associated with the indexes $i_1, i_2... i_v$ are said to be "*extremely close*" when all their sub-neighbourhood of v elements have the same elements. ∎
This is denoted as $[i_1, i_2... i_v] \Leftrightarrow \Gamma_1^v = \Gamma_2^v = \Gamma_v^v$.

Lemma 2 - In each set of points there is at least one pair of them that are extremely close.
Proof. Let $d_{i*j*}$ be the smallest non-zero element in the ($d_{ij}$) matrix, then $j^*$ belongs to $\Gamma_{i*}$ due to the presence of at least two elements in each neighbourhood. In other words, $d_{i*j*} = \text{Min}_j d_{i*j}$, and $d_{i*j*} = \text{Min}_i d_{ij*}$, so that $i^*$ belongs to $\Gamma_{j*}$ due to the symmetry of the matrix and lemma 1. Consequently $\Gamma_{i*}^2 = \Gamma_{j*}^2$, hence $[i^*, j^*]$ are extremely close. ∎

In each step of the tree generation, only the extremely close points are clustered. The new "point" formed by merging the extremely close points is called a pseudo-point and its parameters are taken as the mean of the clustered point parameters. The idea behind such a procedure is that the pseudo-points built in this way are homogeneous. In other words, given two absolutely close points $i$ and $i'$ belonging to $[i_1, i_2,..., i_v]$, then $\forall j \notin [i_1, i_2,..., i_v]$, it is true that $d_{ii'} \leq d_{ij}$ and $d_{ii'} \leq d_{i'j}$. This means that two points in the same cluster have a greater similarity to each other than to any one outside the cluster. The procedure iterates using the new pseudo-points and the remaining substituents, until only one pseudo-point is left. This way, several cluster levels are obtained. The following algorithm might accomplish the generation of the cluster tree:

*2.1 An Algorithm for the Cluster Tree Generation*

Step 1. Compute the normalised parameters.
Step 2. Compute the distance matrix and the cut-off distance.
Step 3. Determine the points' neighbourhoods and identify "extremely close" sets.
Step 4. Merge extremely close points generating pseudo-points which parameters are the mean-value of the clustered point parameters.
Step 5. Stop if a single pseudo-point is found. Otherwise return to step 2.

The algorithm can cluster *at a single step several pairs, triplets' etc*. Trees derived from the modified algorithm are therefore compact and less complex than stepwise hierarchical clustering. A naive illustrative example is presented in the sequel.

## 3 An Application to Substituents in Chemical Compounds

Suppose that Hansch hydrophobic constant π and the Hammet constant σ are the two parameters correlated with the activity of a given chemical series [Hansch and Leo, 1980]. An appropriate subset of substituents can be selected by cluster techniques [Hansch and Unger, 1973; Wooton et al., 1975]. The π and σ parameter values are taken from 25 substituent candidates [Hansch and Leo, 1980] at both *para* and *meta* site (*table 1*). Cluster trees are built by the procedure shown therein. The *table 2* shows the several steps required for obtaining dendrograms as well as the points and pseudo-points. Distance matrices are calculated using normalised parameters, followed by the computation of the cut-off distance. The sub-neighbourhoods are shown on *table 2*. In the *para* position case, the points labelled {Cl and Br}, {Me and $CH_2Me$}, {F and H}, {$SO_2Me$ and $NO_2$}, {$(CH_2)_2Me$ and $CHMe_2$}, {$(CH_2)_3Me$ and $CMe_3$}, {$(CH_2)_6Me$ and $(CH_2)_7Me$} as well as {$(CH_2)_8Me$ and $(CH_2)_9Me$} are extremely close and therefore are clustered (in a single step). After clustering these substituents, the new parameter mean values are calculated and the procedure iterates until only one pseudo-point is achieved. Data concerning *para* and *meta* substitution clustering and dendrograms are shown on F*igure 1* and *2*, respectively.

Clustering approaches have also been combined with Quantitative Structure-Activity Relationships (Q.S.A.R.) in the framework of drug design so as to search for new potentially active drugs [Santos Magalhães et al., 1999].

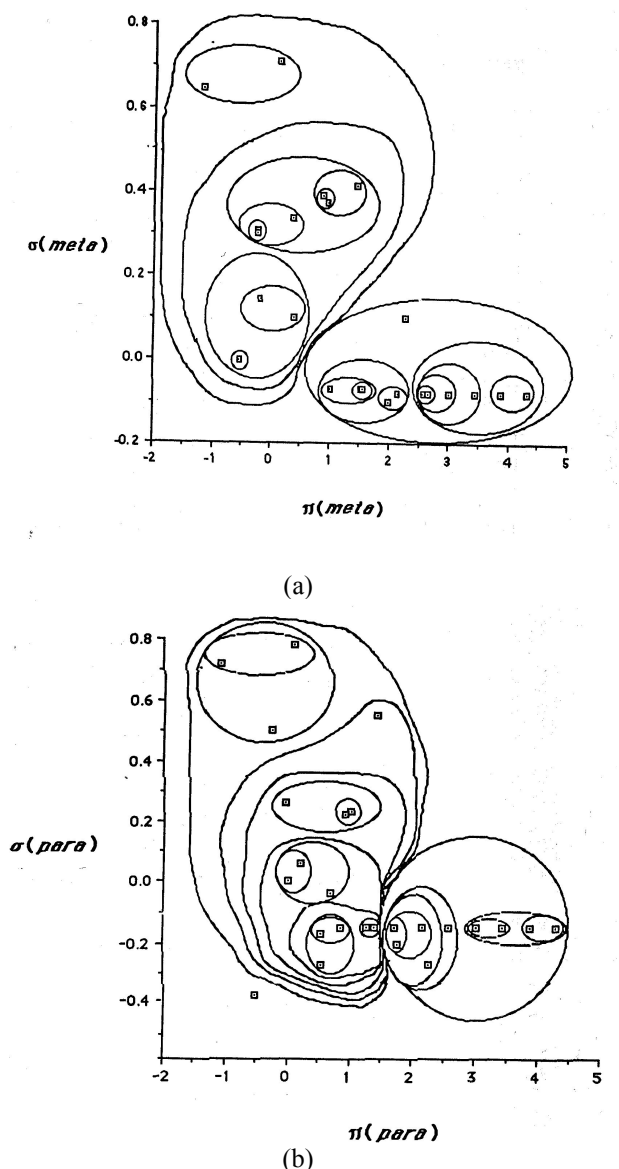

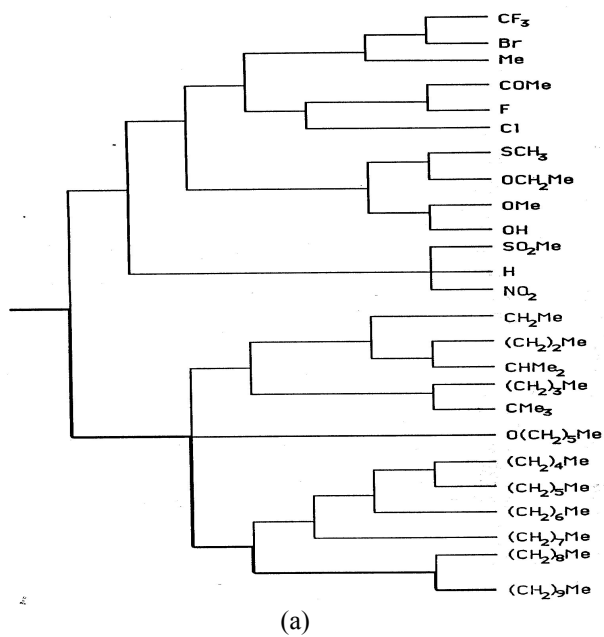

(a)

Figure 1. Clustering sets at several depths for (a) *meta* and (b) *para* positions.

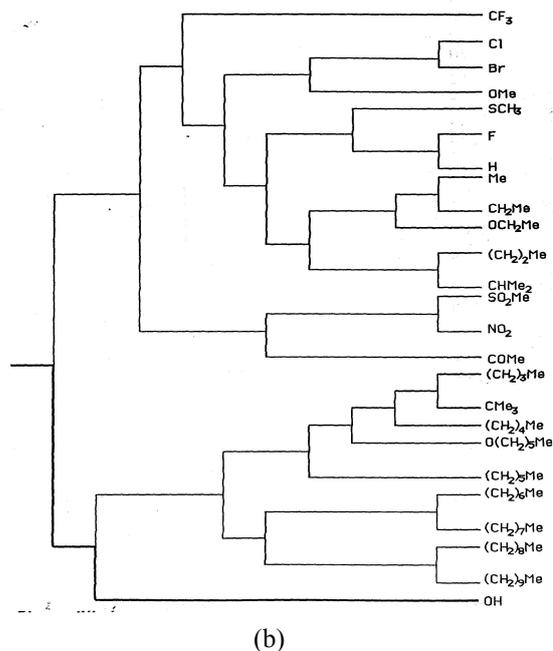

(b)
Figure 2. Cluster tree (dendrogram) for *meta* (a) and *para* (b) positions.

## 4  Conclusions

In this paper a variant of the classical hierarchical cluster analysis is derived. A naive illustrative example showed how to generate cluster trees based on the adaptive algorithm. The superiority of the adaptive clustering with respect to the average-linkage algorithm follows because it achieves a good compromise on threshold values. *Thresholds based on the cut-off distance are sufficiently small to assure the homogeneity and also large enough to guarantee at least a pair of merging sets.*

Trees derived from the modified algorithm are more compact than those ones from stepwise hierarchical clustering. Trees can be built off-line, once and for all. Furthermore, the tree generation depends only on descriptor parameters. If the tree is based on a certain set of descriptors, it can be stored for later use in any related problem with the same descriptors. This approach does not concern directly the problem of collinearity [Wonnacott and Wonnacott, 1990] but correlation matrices derived from sample covariance can be considered in place of the distance matrices.

### Acknowledgements

The author thanks Dr. N.S. Santos Magalhães for her constructive criticism. He is also grateful to Dr. Socrates C.H. Cavalcanti for suggesting the chemical substituent set. Thanks to an anonymous referee for interesting comments.

Table 1. Set of possible substituents at *para* and *meta* positions.

| n. | Substituent | $\pi_p$ | $\sigma_p$ | $\pi_m$ | $\sigma_m$ |
|---|---|---|---|---|---|
| 01 | $CF_3$ | 1.42 | 0.55 | 0.95 | 0.373 |
| 02 | Cl | 0.91 | 0.22 | 0.37 | 0.337 |
| 03 | SMe | 0.69 | -0.04 | -0.20 | 0.144 |
| 04 | Me | 0.55 | -0.17 | 1.45 | 0.415 |
| 05 | F | 0.20 | 0.06 | -0.27 | 0.306 |
| 06 | OMe | -0.05 | 0.26 | -0.54 | -0.002 |
| 07 | $SO_2Me$ | -1.06 | 0.72 | -1.20 | 0.647 |
| 08 | H | 0.00 | 0.00 | 0.10 | 0.710 |
| 09 | $CH_2Me$ | 0.86 | -0.15 | 1.02 | -0.070 |
| 10 | $(CH_2)_2Me$ | 1.29 | -0.15 | 1.55 | -0.070 |
| 11 | $(CH_2)_3Me$ | 1.72 | -0.15 | 2.13 | -0.080 |
| 12 | $(CH_2)_4Me$ | 2.15 | -0.15 | 2.67 | -0.080 |
| 13 | $(CH_2)_5Me$ | 2.58 | -0.15 | 2.58 | -0.080 |
| 14 | $(CH_2)_6Me$ | 3.01 | -0.15 | 3.01 | -0.080 |
| 15 | $(CH_2)_7Me$ | 3.44 | -0.15 | 3.44 | -0.080 |
| 16 | $(CH_2)_8Me$ | 3.87 | -0.15 | 3.87 | -0.080 |
| 17 | $(CH_2)_9Me$ | 4.30 | -0.15 | 4.30 | -0.080 |
| 18 | $CHMe_2$ | 1.40 | -0.15 | 1.53 | -0.070 |
| 19 | $CMe_3$ | 1.78 | -0.20 | 1.98 | -0.100 |
| 20 | $OCH_2Me$ | 0.54 | -0.27 | 0.38 | 0.100 |
| 21 | $O(CH_2)_5Me$ | 2.26 | -0.27 | 2.26 | 0.100 |
| 22 | Br | 1.01 | 0.23 | 0.86 | 0.390 |
| 23 | COMe | -0.27 | 0.50 | -0.27 | 0.300 |
| 24 | OH | -0.54 | -0.37 | -0.54 | -0.002 |
| 25 | $NO_2$ | 0.10 | 0.78 | 0.10 | 0.710 |

Table 2. Adaptive clustering steps.

| depth | cut-off (*para*) | cut-off (*meta*) | Absolutely close points (*para*) | Absolutely close points (*meta*) |
|---|---|---|---|---|
| 1 | 1.05 | 0.89 | [2,22] [4,9] [5,8] [7,25] [10,18] [11,19] [14,15] [16,17] | [1,22] [3,20] [5,23] [6,24] [7,8,25] [10,18] [11,19] [12,13] [16,17] |
| 2 | 1.05 | 1.50 | [4,14] [10,9] | [1,4] [3,6] [8,9] [11,12] |
| 3 | 1.02 | 1.42 | [3,5] [9,13] | [2,4] [8,9] |
| 4 | 0.97 | 1.36 | [2,5] [4,7] [8,9] | [1,2] [5,6] [7,8] |
| 5 | 1.07 | 1.26 | [3,4] [5,9] [7,8] | [1,2] [4,5,6] |
| 6 | 1.13 | 1.65 | [2,3] [5,6] | [1,2] |
| 7 | 1.75 | 2.00 | [1,2] | [1,2] |
| 8 | 1.77 | | [1,2] | |
| 9 | 1.97 | | [2,3] | |
| 10 | 2.00 | | [1,2] | |